\documentclass[sigconf,authorversion]{acmart}
\usepackage{listings}
\usepackage{wrapfig}
\usepackage{tikz}
\usepackage[labelfont=md,textfont=md]{caption}
\usepackage{booktabs}

\newcommand{\bugnum}{27}
\newcommand{\logicbugnum}{19}

\newcommand{\fixedbugnum}{20}
\newcommand{\confirmedbugnum}{25}
\newcommand{\optimizationbugnum}{8}

\usepackage{array}
\newcolumntype{L}[1]{>{\raggedright\let\newline\\\arraybackslash\hspace{0pt}}m{#1}}
\newcolumntype{C}[1]{>{\centering\let\newline\\\arraybackslash\hspace{0pt}}m{#1}}
\newcolumntype{R}[1]{>{\raggedleft\let\newline\\\arraybackslash\hspace{0pt}}m{#1}}
\usepackage{amsmath}
\usepackage{algorithm}
\usepackage[noend]{algpseudocode}
\usepackage{multirow}
\usepackage{balance}
\usepackage{listings}
\usepackage{color}
\definecolor{dkgreen}{rgb}{0,0.6,0}
\definecolor{gray}{rgb}{0.5,0.5,0.5}
\definecolor{mauve}{rgb}{0.58,0,0.82}
\usepackage[clean]{revdiff}
\makeatletter
\def\BState{\State\hskip-\ALG@thistlm}
\makeatother
\algnewcommand\algorithmicswitch{\textbf{switch}}
\algnewcommand\algorithmicdefault{\textbf{default:}}
\algnewcommand\algorithmiccase{\textbf{case}}
\algnewcommand\algorithmicassert{\texttt{assert}}
\algnewcommand\Assert[1]{\State \algorithmicassert(#1)}%
\algdef{SE}[SWITCH]{Switch}{EndSwitch}[1]{\algorithmicswitch\ #1\ \algorithmicdo}{\algorithmicend\ \algorithmicswitch}%
\algdef{SE}[CASE]{Case}{EndCase}[1]{\algorithmiccase\ #1}{\algorithmicend\ \algorithmiccase}%
\algdef{SE}[DEFAULT]{Default}{EndDefault}[1]{\algorithmicdefault\ #1}{\algorithmicend\ \algorithmicdefault}%
\algtext*{EndDefault}%

\usepackage{xcolor}
\usepackage{lipsum}
\newcommand{\hli}[1]{%
  \colorbox{lightgray!45}{{#1}}%
  }

\acmConference[ICSE 2024]{46th International Conference on Software Engineering}{April 2024}{Lisbon, Portugal}
\AtBeginDocument{%
  \providecommand\BibTeX{{%
    \normalfont B\kern-0.5em{\scshape i\kern-0.25em b}\kern-0.8em\TeX}}}

\copyrightyear{2024}
\acmYear{2024}
\setcopyright{rightsretained}
\acmConference[ICSE '24]{2024 IEEE/ACM 46th International Conference on Software Engineering}{April 14--20, 2024}{Lisbon, Portugal}
\acmBooktitle{2024 IEEE/ACM 46th International Conference on Software Engineering (ICSE '24), April 14--20, 2024, Lisbon, Portugal}\acmDOI{10.1145/3597503.3639208  }
\acmISBN{979-8-4007-0217-4/  24/04}

\begin{document}

\title[Finding XPath Bugs in XML Document Processors via Differential Testing]{Finding XPath Bugs in XML Document Processors \\ via Differential Testing}

\author{Shuxin Li} 
\authornote{Work done during an internship at the National University of Singapore.}
\orcid{0009-0003-0468-2029}
\email{shuxin.li.lv@gmail.com}
\affiliation{%
  \institution{Southern University of Science and Technology}
  \country{China}
}

\author{Manuel Rigger}
\email{rigger@nus.edu.sg}
\orcid{0000-0001-8303-2099}
\affiliation{%
  \institution{National University of Singapore}
  \country{Singapore}}



\begin{abstract}
Extensible Markup Language (XML) is a widely used file format for data storage and transmission. Many XML processors support XPath, a query language that enables the extraction of elements from XML documents. These systems can be affected by logic bugs, which are bugs that cause the processor to return incorrect results. In order to tackle such bugs, we propose a new approach, which we realized as a system called XPress. As a test oracle, XPress relies on differential testing, which compares the results of multiple systems on the same test input, and identifies bugs through discrepancies in their outputs. As test inputs, XPress generates both XML documents and XPath queries. Aiming to generate meaningful queries that compute non-empty results, XPress selects a so-called targeted node to guide the XPath expression generation process. Using the targeted node, XPress generates XPath expressions that reference existing context related to the targeted node, such as its tag name and attributes, while also guaranteeing that a predicate evaluates to true before further expanding the query. We tested our approach on six mature XML processors, BaseX, eXist-DB, Saxon, PostgreSQL, libXML2, and a commercial database system. In total, we have found \bugnum{} unique bugs in these systems, of which \confirmedbugnum{} have been verified by the developers, and \fixedbugnum{} of which have been fixed. XPress is efficient, as it finds 12 unique bugs in BaseX in 24 hours, which is 2$\times$ as fast as naive random generation. We expect that the effectiveness and simplicity of our approach will help to improve the robustness of many XML processors.
\end{abstract}



\maketitle

\section{Introduction}
Extensible Markup Language (XML) is a widely used file format for data storage and transmission. XPath is an expression language, which provides the ability to navigate through XML documents to select wanted nodes. XPath is also at the core of other XML query language standards such as XSLT~\cite{XSLT-ref} and XQuery~\cite{XQuery-ref}, making it a fundamental XML query language. 

Various XML document processors have been developed for extracting information from XML documents efficiently. We loosely categorize them depending on whether they can store XML documents in addition to processing them---that is, whether they are Database Management Systems (DBMSs), or provide only processing functionality.
In terms of DBMSs specialized for XML, popular examples include BaseX~\cite{BaseX} and eXist-DB~\cite{eXist-DB}.
Many general-purpose DBMSs such as Oracle Database~\cite{Oracle}, MySQL~\cite{MySQL}, and PostgreSQL~\cite{PostgreSQL} have adopted support for processing XML documents. In fact, out of the 10 most popular DBMSs according to the DB-engines ranking~\cite{DBEngines}, 6 support at least partial XML document parsing.
A popular example of a processor without storage functionality is Saxon.
Saxon~\cite{Saxon} is an in-memory XML processor that can be either used in a standalone way or embedded as a library.
Finally, libxml2~\cite{libXML2} is a popular XML processing library written in C.
XPath is supported by all of these processors.

XML document processors can be affected by logic bugs. Logic bugs are bugs that cause the XML processor to produce incorrect results without crashing the system, meaning that they can often go unnoticed.
In order to find such bugs, developers mainly rely on test suites such as the XPathMark test suite for XPath~\cite{Franceschet05}, the W3C qt3 test suite~\cite{W3Cqt3}, or hand-written unit tests.
Manually writing tests requires much effort, and it is challenging to comprehensively cover the XML processors' functionality.
To find logic bugs automatically, a so-called \emph{test oracle} is required that can determine whether the system's output is correct in order to find logic bugs.
Todic and Uzelac have proposed an automated testing technique for SQLServer's index support; their test oracle compared the results of a given query with and without index definition~\cite{Todic12}. A limitation of this technique is that it is applicable only to finding index-related bugs in DBMSs. To the best of our knowledge, no other test oracles have been proposed in this context.

In order to detect XPath-related bugs in XML processors, we propose differential testing as an oracle. The core idea of differential testing is to use one input that is executed using multiple systems; any discrepancy in the results indicates a potential bug in the system. For testing XML processors, the input for the XML processors under test is an XML document and XPath expression, and the results are a sequence of XML nodes or values. Differential testing has been successfully applied in various related domains, such as relational DBMSs~\cite{slutz1998massive}, compilers~\cite{yang2011finding, zhang2017skeletal}, JVM implementations~\cite{Chen16}, ORM systems~\cite{Sotiropoulos21}, and graph DBMSs~\cite{Zheng22, Hua23}. Its effectiveness hinges on two main requirements. First, multiple systems to be compared must be available. As discussed above, various XML processors with XPath support exist. Second, for any valid input, the systems should produce the same result, since otherwise, a differential-testing approach raises many false alarms. This requirement is not always met, for example, when applying differential testing to relational DBMSs, where the \emph{``common SQL subset is relatively small and changes with each release"} and \texttt{NULL} handling differs between DBMSs~\cite{slutz1998massive}. As we found, XPath is a well-defined language by the W3C standard, and XPath implementations of the same standard follow the same language rules, making differential testing highly applicable.

To generate test cases, we propose an approach that selects a so-called \emph{targeted node} from the XML document, based on which we generate a query that is guaranteed to fetch at least that node.
As such, it tackles two challenges that might prevent testing from exercising interesting behaviors. First, by generating the query based on the targeted node, we can guarantee that we access a tag name, attributes, and relative paths that exist with respect to at least the targeted node. Second, by rectifying predicates so that they evaluate to true for the targeted node, we can ensure that the result set is non-empty even for complex queries.
A similar high-level idea has been proposed in the context of testing relational DBMSs, called \emph{Pivoted Query Synthesis (PQS)}~\cite{rigger2020testing}, where a pivot row was selected, based on which predicates were rectified to return true.
Apart from applying that idea in a different context, we also propose a different rectification strategy that eschews mirroring the predicate's execution logic in the testing tool, which was required for realizing PQS.

We implemented our approach as a tool named XPress,\footnote{Our artifact is publicly available at \url{https://zenodo.org/records/10473926}} which, to the best of our knowledge, is the first \emph{general} automated testing tool for XML processors, and tested our method on six mature and widely-used XML processors BaseX, eXist-DB, Saxon, PostgreSQL, libXML2 and a commercial DBMS.
The experimental results show that our approach is effective in detecting XPath-related logic bugs in XML processors. We found \bugnum{} previously unknown unique bugs, not covered by existing test suites, of which \logicbugnum{} were logic bugs. \confirmedbugnum{} of them have been confirmed, and \fixedbugnum{} of them have been fixed. Furthermore, these test cases have been integrated into the aforementioned qt3 test suite, so that they can detect potential bugs in XML processors that we have not tested. Our experiments demonstrate that our proposed guided query generation process improves testing efficiency by finding 2$\times$ more unique bugs within 24 hours in BaseX as compared to random generation. Given the high effectiveness and efficiency of the approach, we believe it will likely be adopted by developers of XML processors to improve their systems.

To summarize, we make the following contributions: 
\begin{itemize}
  \item We propose the first general approach for automatically testing XML processors in order to find logic bugs.
  \item We implemented and evaluated our approach on six widely-used XML process systems, which successfully found \bugnum{} previously-unknown unique bugs. 
\end{itemize}

\section{Background} \label{background}

\begin{figure}[bt]
  \centering
  \includegraphics[width=\linewidth]{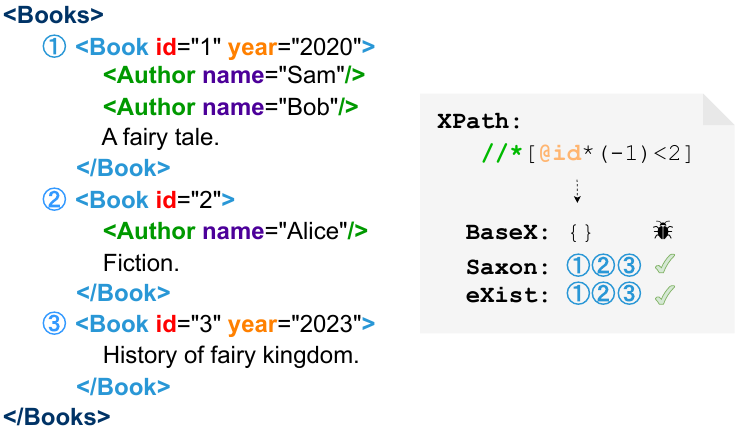}
  \caption{Example XML and motivating example.}
  \label{fig:example}
\end{figure}

\paragraph{Running example}
 Figure~\ref{fig:example} shows a running example that we will subsequently use to explain basic XML and XPath concepts and outline the challenges of automated testing as applied in this context.
 The left shows an XML document with the root node \hli{Books}, while the right shows an XPath expression \hli{//*[@id * -1 < 2]}.
 We adapted this example from a bug-inducing test case that XPress discovered.\footnote{\url{https://github.com/BaseXdb/basex/issues/2188}}
 As shown, for the query on the document, BaseX returned an empty result, while both Saxon and eXist returned all three \hli{Book} nodes.


\paragraph{XML} Extensible Markup Language (XML) is a text format for describing structured data. XML documents are trees that consist of nodes, as illustrated in Figure \ref{fig:example}. An XML document has a single root element node (see \hli{<Books>}). Each element node has a tag name (see \hli{Books}, \hli{Book}, and \hli{Author}). Element nodes can include attribute nodes. For example, two of the \hli{<Book>} nodes have both attribute nodes \hli{id} and \hli{year}. An element node can also include child element nodes; in the example, the \hli{<Books>} node contains three child element nodes \hli{<Book>}. Element nodes can hold \emph{text contents}, which can be of any defined data type. 
For the \hli{<Book>} node with attribute \hli{id = 1}, the text content it holds is \hli{"A fairy tale"}. Attribute nodes are disallowed from holding child nodes. In the example, \hli{id} and \hli{year} are integer-typed attribute nodes and the \hli{name} attributes are string-typed attribute nodes.  


\paragraph{XPath.} The XPath language is an expression language that allows navigating the XML tree and hierarchic addressing of the element nodes. XPath is at the core of both \emph{eXtensible Stylesheet Language - Transformation (XSLT)}~\cite{XSLT-ref} and \emph{XQuery}, a more expressive query language for XML~\cite{XQuery-ref}. XSLT transforms XML documents into other formats and the XQuery language is a super-set of XPath expressions. XQuery extends XPath to provide functionalities such as node constructors and SQL-like clauses. 

\begin{figure}[bt]
  \centering
  \includegraphics[width=\linewidth]{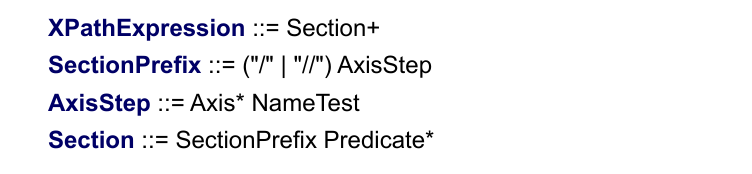}
  \caption{Simplified structure of XPath expressions.}
  \label{fig:structure}
\end{figure}

\paragraph{XPath structure.} An XPath expression describes the selection and transformation of nodes of the XML tree. Figure~\ref{fig:structure} shows a simplified XPath 3.0~\cite{XPath-3.0-ref} grammar using EBNF notation from the W3C XML 1.0 standard~\cite{ebnf-notation}. We introduce the non-established terms \textit{Section} and \textit{Section Prefix} to describe our generation approach in Section~\ref{xpath}. XPath expressions consist of one or more sections, and a section contains one section prefix followed by zero or more predicates. In Figure~\ref{fig:example}, the XPath expression \hli{//*[@id*(-1)<2]} consists of a single section with section prefix \hli{//*} and a single predicate \hli{[@id*(-1)<2]}. Each section prefix starts with either \hli{/} or \hli{//}. \hli{/} is called the \emph{path operator}, which accepts a node sequence as the left-hand operand and orders it in document order while eliminating duplicate nodes. \hli{//} represents the abbreviated relative path \hli{/descendant-or-self::node()/}, which matches the current context and all descendant nodes of the current context, regardless of the intermediate path. An axis step consists of an optional axis and a name test. 

\paragraph{XPath axes.} Axes define the relationship between selected nodes and current context nodes. For example, the axis \hli{parent::} selects all parent nodes of current context nodes. If omitted, it is equivalent to \hli{child::}, which selects all direct children nodes of current context nodes. A name test is a string literal to fetch only nodes with the same tag name. It could also be a wildcard \hli{*}, which matches all nodes without applying filters. The section prefix \hli{//*} in the example selects all descendant nodes of the document node, which is all element nodes in the document. 

\paragraph{XPath predicates.} Predicates in XPath include positional predicates and boolean predicates. Positional predicates contain an expression that evaluates to a single integer and select only values whose position in the context matches the integer value. In the XPath expression \hli{/Books/Book[1]}, \hli{[1]} is a positional predicate and selects only the first child of \hli{<Books>}, which is the \hli{<Book>} node with \hli{@id=1}. Boolean predicates evaluate current context nodes to a boolean value according to a given expression and only nodes for which the predicate evaluates to \hli{true} are selected. In Figure \ref{fig:example}, \hli{[@id * -1 < 2]} is a boolean predicate. The query \hli{//*[@id * -1 < 2]} selects all nodes in the XML document with attribute \hli{id} that satisfy \hli{id * -1 < 2}. The three nodes with tag name \hli{Book} in the document have attribute \hli{id}, and all satisfy the condition. Therefore, if correctly evaluated, this query should return all three \hli{Book} nodes. 

\paragraph{Logic bug.} For the test input in Figure~\ref{fig:example}, systems like Saxon and eXist-DB both returned a result set with three \hli{Book} nodes, while BaseX returned an empty result set. The difference between the processors indicates a potential bug. Based on our manual analysis, we suspected that BaseX computed an incorrect result, which is why we reported it to the BaseX developers. They fixed the bug quickly. The reason for this bug was an incorrect simplification of the arithmetic expression \hli{x * a > b} to \hli{x > b / a}. When the divisor is a negative number, the original operator \hli{>} should be reversed to \hli{<}. 

\paragraph{XPath standard.} There are majorly two different standards of XPath implementations in use today, which we need to consider in our work. The XPath 1.0 standard was the first version. As a super-set of XPath 1.0, the XPath 3.0 standard is the latest standard of the XPath language and provides more functionalities such as advanced data types and functions~\cite{XPath-3.0-functions-ref}. Most multi-model DBMSs, which support XPath queries, support only XPath 1.0~\cite{XPath-1.0-ref} (\textit{e.g.}, Oracle, MySQL, and PostgreSQL). While some specialized XML processors support also only XPath 1.0 (\textit{e.g.}, libXML2), others support the more recent XPath 3.0 standard (\textit{e.g.}, BaseX, eXist-DB, and Saxon).

\paragraph{XPath versions and differential testing}
The same queries might produce different results under different standards. For example, for the XPath expression \hli{Book/@name = false()}, under the XPath 1.0 standard, the expression is expected to return true. \hli{@name} is first cast into its equivalent boolean value. In the current case \hli{<Book>} has no \hli{name} attribute, therefore, an empty node set is returned. The equivalent boolean value evaluates to false for empty nodes. Comparing false to false is equal, therefore true is returned. Under the XPath 3.0 standard, however, the result is expected to be false. \hli{@name} returns an empty sequence and equality comparison between an empty sequence and a boolean value false would evaluate to false.
Thus, applying differential testing to XML processors supporting different standards is infeasible.

\begin{figure*}
  \includegraphics[width=\textwidth]{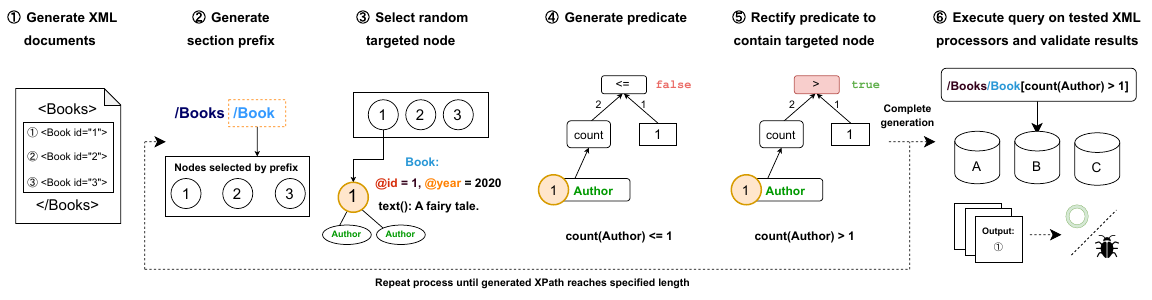}
  \caption{Overview of the approach implemented in XPress.}
  \label{fig:overview}
\end{figure*}

\section{Approach} 
Figure~\ref{fig:overview} shows an overview of the approach using the same example as in Figure~\ref{fig:example}. At a high level, our approach consists of three main steps. First, we randomly generate an XML document as the context for the following queries (step \textcircled{1}). We then generate an XPath expression that we will subsequently validate (step \textcircled{2} to step \textcircled{5}). Finally, we execute the XPath expression on the XML document using all engines under test and compare the resulting outputs to detect potential bugs (step \textcircled{6}). In the subsequent subsections, we explain these steps in reverse order to reflect their importance.

We guide the XPath expression generation towards queries that reference nodes and attributes present in the XML document and result in non-empty result sets based on the intuition that they are more likely to stress the underlying logic of the tested systems. To generate XPath expressions with non-empty result sets, we construct the query section-by-section and ensure that a non-empty result set is produced before proceeding with the next section. Each section consists of a section prefix and predicates, and we first generate the prefix (step \textcircled{2}) and then the predicate. By restricting the section prefix, we guarantee that the result contains at least one node. From the nodes selected by the section prefix, we randomly select a node as a target (step \textcircled{3}). We then generate a predicate aiming to select the targeted node using a bottom-up tree construction method (step \textcircled{4}). We rectify the predicate to ensure that the result set contains the targeted node (step \textcircled{5}). We repeat this process until the XPath query reaches the desired length.

\subsection{Differential Testing for XML Processors} \label{differential in approach}
As detailed subsequently, differential testing enables us to find both logic bugs as well as internal errors when comparing the results of XML processors implementing the same XPath standard.

\paragraph{Query execution} When passing XML documents and XPath queries to different systems, we must account for the different input interfaces. For example, DBMSs use database connection interfaces to store and query data, while Saxon can be used as a library. To abstract this, we treat every XML processor implementation as a function that returns a result set and expects two string values, namely an XML document \lstinline{XML} and an XPath query \lstinline{XPATH}. Listing~\ref{lst:oracledb} shows \rold{how our Oracle Database component implements this interface using SQL statements.}\rnew{an implementation of this interface for Oracle Database using SQL statements.} It creates a table \lstinline{t}, inserts the XML document---the \lstinline{XMLType} constructor is used to convert the string to an XML data type---and uses an \lstinline{XMLQuery} function call in a \lstinline{SELECT} statement to compute the result set. 
\rnew{For BaseX and eXist-db, similar to the commands shown for Oracle Database, we also start with an empty database and subsequently insert an XML document.}
Listing~\ref{lst:saxon} shows an excerpt of the Java code for Saxon. First, the call to \lstinline{compile} converts the textual XPath query to an executable object, which is then loaded.  Unlike\excludecomment{for Oracle Database, which requires inserting data into the database, }\rnew{for the DBMSs, which require inserting data into a database, }for Saxon, the XML document is simply associated with the query using the \lstinline{setContextItem} call. The \lstinline{evaluate} call computes the result, which is returned for comparison. 

\begin{figure}[tb]
\begin{lstlisting}[caption={Execution of XPath using Oracle Database}, mathescape=true, label={lst:oracledb}]
$\textbf{CREATE TABLE}$ t (a XMLType);
$\textbf{INSERT INTO}$ t $\textbf{VALUES}$ (XMLType($\underbar{XML}$)));
$\textbf{SELECT XMLQuery}$($\underbar{XPATH}$ PASSING a RETURNING CONTENT) $\textbf{FROM}$ t;
\end{lstlisting}
\vspace{-7mm}
\end{figure}

\begin{figure}[tb]
\begin{lstlisting}[caption={Execution of XPath using Saxon in Java}, mathescape=true, label={lst:saxon}]
XQueryExecutable exec = compiler.compile($\underbar{XPATH}$);
XQueryEvaluator query = exec.load();
query.setContextItem($\underbar{XML}$);
XdmValue result = query.evaluate();
\end{lstlisting}
\vspace{-5mm}
\end{figure}

\rnew{\paragraph{Bug identification.} We identify both logic bugs and internal errors by comparing the returned results of different processors on the same XML document and query}.
We identify logic bugs when the tested systems return different node-set outputs for the same test cases. To parse and track the results easily under different output formats, we use unique node ids to identify element nodes. 
We detect internal errors as discrepancies with respect to errors. Rather than checking for an exact error message match, we validate whether all the systems produce an error, or all execute the XPath query successfully. \rnew{If only a subset of the systems report an error for the same XPath query, we found a potential bug.}


\paragraph{Different XPath standards.} Our approach and tool are applicable to both XPath 1.0 and XPath 3.0. However, due to the differences in the formats, only processors using the same standard can be tested. Functionality that is supported only in XPath 3.0, can be disabled while generating test cases for processors that implement XPath 1.0. 
For example, sequence functions, such as \hli{subsequence}, are defined only for the XPath 3.0 standard. 
When generating test cases for XPath 1.0 processors, we omit to generate \hli{subsequence} function nodes for predicates. We did not encounter any functions or operators that were removed in the XPath 3.0 standard, so all expressions that we generate when testing XPath 1.0 processors can be used also when testing XPath 3.0 processors. By comparing only processors with the same XPath standard against each other, the difference in the results between different XPath standards (see Section \ref{background}) has no influence on the testing process.

\subsection{XPath Expression Generation} \label{xpath}
In this section, we introduce how we generate XPath queries. We encountered two main challenges that we had to tackle when generating XPath expressions. 

\paragraph{Non-existent elements} Randomly generated queries could be semantically correct, but reference non-existent nodes or attributes. For the document in Figure~\ref{fig:example}, \hli{//Author[@id < 1]} is a valid XPath expression. However, none of the \hli{Author} nodes contain an \hli{id} attribute. Thus, XPath returns an empty sequence for each node, causing the predicate \hli{@id < 1} to evaluate to false. We believe that queries, where only non-existent attributes or nodes are referenced, are less likely to exercise the logic of the processors under test, as subsequent operations are likely to evaluate to an empty sequence as well. Thus, we aim to avoid generating such queries.

\paragraph{Empty results} Randomly generated predicates might likely evaluate to false and cause queries to generate empty result sets. For the document in Figure~\ref{fig:example}, the XPath predicate \hli{starts-with(text(), x)} identifies nodes whose text starts with \textit{x}. If \textit{x} is a randomly generated string, the possibility is high that no nodes in the current result set match the condition. Consequently, any use of the predicate would yield an empty result. 
Any subsequently added section would yield an empty result as well, meaning that such queries would be less likely to exercise the processor under test. Consequently, we want to avoid generating such predicates, in particular, when they involve multiple sections. This relates to the first problem, as non-existent nodes or attributes can also introduce empty results.

\paragraph{Approach overview}
We designed the XPath generation process of XPress tackling the two aforementioned issues. To create XPath expressions that refer to valid nodes and attributes to trigger deeper logic of the system under test, we generate queries that reference existent context relative to the so-called targeted node, such as its tag name and attributes (steps \textcircled{3} and \textcircled{4}). Since randomly generated predicates might miss the targeted node from the result set, we rectify the generated expressions to ensure the inclusion of the targeted node (step \textcircled{5}). 

\paragraph{Iterative section generation} We create XPath expressions section-by-section by executing step \textcircled{2} to step \textcircled{5} for each section, which allows us to ensure non-empty results after generating each section. 
In the example, we first generate section \hli{/Books} by selecting \hli{<Books>} as the targeted node, and after executing steps \textcircled{2} to step \textcircled{5}, the result of \hli{/Books} is non-empty---containing the node \hli{<Books>}. Based on this, we further proceed to generate the next section \hli{/Book[count(Author) > 1]} starting at step \textcircled{2}. 

\paragraph{Section prefix} We randomly generate one of the applicable section prefixes. First, we randomly select the start of the section to be \hli{/} or \hli{//}. We then retrieve the current context node sequence by executing the expression---\hli{/Books} in the example---on a processor. Based on the result, we include all possible axes that would not lead to an empty result set by simple conditional checks. We support all 11 axes described in the XPath 3.0 standard~\cite{XPath-3.0-ref}. For example, applying the axis \hli{/descendants::} will lead to a non-empty result, if at least one non-leaf node exists in the current selection. From the possible axes, we select a random one and apply it. When generating the section prefix \hli{/Book}, the axis step is implicit. It is equivalent to \hli{/Books/child::Book}, which selects all child nodes of the previously selected nodes. We again execute the query and retrieve the result node-set. We use the result for the name test, for which we either select a tag name from the result node-set, or use the wildcard \hli{*}. By doing so, we are again guaranteed a non-empty result set. In the example, the tag name \hli{Book} is selected and applied, resulting in the selection of all three \hli{Book} nodes.
In our artifact, we include a table that details the conditional checks for all 11 axes.

\paragraph{Target node selection} To generate targeted queries that fetch at least one node, we select a so-called \emph{targeted node} to guide the predicate generation process. We use information about the target node, such as its text content, the attributes it holds, and its relationship to other nodes during the predicate generation. This is similar to the concept of the pivot row in PQS~\cite{rigger2020testing}, which is a technique that has been proposed to test relational DBMSs. After the generated predicate is applied, we expect the target node to be included in the result node-set. In step \textcircled{3}, we select node 1 as the target node for the predicate generation process. Constraining the context to exist for the targeted node does not affect the evaluation of the expression on other candidate nodes and, therefore, still allows finding bugs that are triggered only when referring to nodes' non-existing attributes or child nodes.

\paragraph{Predicate generation} We use a tree structure to represent the predicate and take a bottom-up construction approach to enable tracking of expression results along tree construction. We start generating the predicate from a specific subject, which is either the targeted node or a node sequence derived from the targeted node with equal probability. In the example, we select the \hli{<Author>} child node sequence from the targeted node as the subject. We then iteratively apply random function nodes and supply function parameters to construct the predicate, until the predicate reaches a desired length. We keep track of the data type and value of the current sub-expression when constructing the predicate, by executing the sub-expression on one randomly chosen XML processor\rnew{---we use this XML processor also for predicate rectification and we subsequently refer to this XML processor as the \emph{designated} XML processor.} We use the value and data type of the current sub-expression in the following two ways: by (1) selecting function nodes of according data types and (2) supplying arguments to reference existent context and triggering corner cases. Specifically, we select a random function node from functions that could accept the value of the current data type as input. A function node can either represent a function or an operation. In the example, \hli{Author} is a node sequence and \hli{count} is a randomly selected function from functions that accept node sequence as input. For function nodes that require additional arguments, we supply arguments while taking the current result value into consideration. As an example, when selecting attribute values from node sequences, we use name tests referencing existent attributes. For the \hli{=} operator, we choose an operand that is equal to the current value with a high probability of triggering the equal case which is of low probability under random generation. Aside from constants, we also set the possibility for operands to be other predicate trees. Through this, we support the generation of expressions with multiple subject occurrences. Besides boolean predicates, we also apply positional predicates to the XPath expression randomly.

\begin{algorithm}[bt]
\caption{Predicate Rectification}\label{predicate rectification algorithm}
\begin{algorithmic}[1]
\Function{RectifyPredicate}{$predicate\_node$, $targeted\_node$}
\State ${c1} \gets predicate\_node.leftChild$
\State ${c2} \gets predicate\_node.rightChild$
\If {targeted\_node in \Call{GetResult}{$predicate\_node$}} \State \Return
\EndIf
\If {\Call{RandomProb}{\null} < 0.5} \State \Call{AddNot}{$predicate\_node$} \State \Return
\EndIf
\Switch{$predicate\_node$}
    \Case{$or\:operator$}
        \If {\Call{RandomProb}{\null} < 0.5} \State \Call{RectifyPredicate}{$c1$}
        \Else \State \Call{RectifyPredicate}{$c2$}
        \EndIf
    \EndCase
    \Case{$and\:operator$}
      \State \Call{RectifyPredicate}{$c1$}
      \State \Call{RectifyPredicate}{$c2$}
    \EndCase
    \Case{$comparison\:operator$}
      \State \Call{ChangeToOpposite}{$predicate\_node$}
    \EndCase
    \Default{}
      \State \Call{AddNot}{$predicate\_node$}
    \EndDefault
\EndSwitch
\State \Return
\EndFunction
\end{algorithmic}
\end{algorithm}

\paragraph{Predicate rectification} Lastly, we rectify the generated predicate to guarantee that the targeted node is contained in the final result set. We first execute the generated predicate \rnew{on the designated XML processor}. If the result set misses the targeted node, we rectify the predicate. To negate the predicate's result, we can always apply a \hli{not} operator. However, as shown in Algorithm \ref{predicate rectification algorithm}, we probabilistically apply more specific rectification for certain operators to uncover additional potential bugs. For logical operators such as \hli{and}, both child expressions need to be modified to evaluate to true to contain the targeted node, while \hli{or} needs only modification of one random child expression. For comparison operators, such as \hli{<=}, we replace them with their opposite operators, which, in the example, is \hli{>}. Thus, the targeted node is guaranteed to be contained in the result set.

\subsection{XML Generation} \label{xml generation}
In this section, we outline how we generate XML documents (step \textcircled{1}), which we do not consider part of our core contribution.


\paragraph{Tree creation.} \rnew{We use a bottom-up approach to generate XML documents. We first generate a number of node templates, which we use to generate XML nodes that have overlaps in terms of structure, as detailed below. We select one of these nodes as a root element. For the remaining nodes, we randomly assign each node to a parent.}\excludecomment{We combine the element nodes into an XML document.}\excludecomment{ We select one node as the root node, and assign a parent to each of the remaining element nodes iteratively.} As XML documents support recursive structure, we allow cyclic relationships. \rnew{In Section 4, we provide details on how we configured the number of nodes in a document.}

\paragraph{Node generation.}\rnew{ We introduce how each element node is instantiated.} By default, XML documents do not have to adhere to a specific schema, which is unlike, for example, relational DBMSs.
Nevertheless, we want to generate element nodes that have overlaps in terms of structure, to test for more interesting behaviors.
To that end, we generate element nodes based on so-called \emph{node templates} that we randomly generate. 
A node template represents a type of node. For example, in Figure \ref{fig:example}, \hli{Book} is a node template whose tag name is \hli{Book}, has attributes \hli{id} and \hli{year}, and has text content of string data type. To instantiate the template, we fill in values for the attributes and text contents.
\rnew{For each node we created in the aforementioned XML tree, we instantiate it with a randomly assigned template. }
In the example of Figure~\ref{fig:example}, we generated three nodes using the \hli{Book} template. We assign random values for element nodes and their attributes according to the associated data types except \hli{id}, to which we assign a unique identifier, which we use to unambiguously identify the processors' outputs (see Section~\ref{differential in approach}). For the \hli{<Book>} node with \hli{id = 1}, we assign the random integer value \hli{2020} to \hli{year} and the random string value \hli{"A fairy tale"} as its text content. Similar strategies have been applied also to other schema-less systems such as graph DBMSs~\cite{kamm2023testing, hua2023gdsmith}.

\section{Evaluation}
In the evaluation, we sought to investigate whether our technique is effective and efficient in finding bugs for XPath expression processors. Specifically, we were interested in the following questions:

\begin{itemize}
    \item [\textbf{Q1.}] Is XPress effective in finding new XPath-related bugs in established XML processors (see Section \ref{effectiveness})? 
    \item [\textbf{Q2.}] Does the query generation approach described in Section \ref{xpath} improve the bug-finding efficiency of XPress with respect to \rnew{real-world baselines and} a\excludecomment{ naive}random generation approach (see Section \ref{efficiency})?
    \item [\textbf{Q3.}] How does the differential testing test oracle compare to the state-of-the-art oracle (see Section \ref{comparison})? 
    \item [\textbf{Q4.}] What kind of XPath-related bugs might be overlooked by XPress (see Section \ref{historical})? 
\end{itemize}

\paragraph{Tested XML Processors.} We tested our method on six mature, well-known, and actively maintained XPath processors: BaseX, exist-DB, Saxon-HE, PostgreSQL, libXML2, and a commercial DBMS, whose name we have omitted due to its ``DeWitt clause''~\cite{Dewitt}. We started testing on BaseX 10.4, eXist-DB 6.2.0, Saxon Home Edition 12.2, PostgreSQL version 15, and libXML2 commit version 106153. As bugs were resolved, we constantly updated to the latest available version. We selected BaseX, eXist-DB, and Saxon to be our main testing targets, because they all implement the more recent XPath 3.0 standard. BaseX ranks as the most popular Native XML DBMS on the DB-Engines Ranking~\cite{DBEngines}.
eXist-DB is widely applied in data centers, systems, and platforms, as referenced on the eXist-DB reference page~\cite{eXist-ref}. Saxon is an in-memory processor and therefore is not included in the DB-Engines rankings. However, the official website of Saxon~\cite{Saxon-ref} states: \emph{"More than 170 software vendors have built Saxon into their own applications"} and \emph{"6 of the world's top 10 software vendors are Saxonica clients"}, demonstrating that Saxon is a widely-used and popular XML processor. For XPath 1.0 standard implementations, we tested PostgreSQL, libXML2, and the commercial DBMS. PostgreSQL is a popular open-source DBMS, which ranks 4 on the DB-Engines ranking and has 12.8k stars on GitHub. libXML2 is a software library developed for the GNOME project. The commercial DBMS is often considered the most popular and important DBMS overall, as also reflected in various rankings. All XML processors have been actively maintained for over 15 years.


\paragraph{Experimental setup.} We implemented the tool, XPress, in around 8,000 LOC in Java. In our experiments, we configured it to generate XML documents that contain 1 to 50 nodes. \rnew{We create half as many node templates as element nodes.} For each XML document, we generated 200 XPath expressions. Each XPath expression had an equal possibility to hold 1 to 7 sections. We set one predicate to hold at most 10 subjects (see Section~\ref{xpath}) and the depth of the predicate tree to be at most 10. \rnew{We used the default settings of each XML processor. } We conducted all our experiments using a personal computer with a 64-Core AMD EPYC 7763 CPU at 2.45GHz and 512GB memory running Ubuntu 22.04.

\subsection{Effectiveness} \label{effectiveness}
In this section, we show XPress' effectiveness through the number of bugs found, developer feedback, and illustrative examples. 

\paragraph{Methodology} \label{effectiveness-methodology}
We implemented the tool while intermittently testing the systems over a period of 3 months.\excludecomment{ For every found discrepancy, we reduced the test case, and if we considered the reduced XPath expression to likely trigger an unknown bug, we reported it to the developers.}\rnew{ For every found discrepancy, we reduced the test case. If the test case exhibited an unreported pattern, we considered it likely to be an unknown bug and reported it to the developers.} Note that this was a best-effort approach, and that it is an unsolved problem of how to identify duplicate bugs effectively. \rnew{Whether we considered a bug as unique was based on the developers' verdict; we considered a bug only as unique if an issue was addressed through an independent bug fix.} Unfixed bugs hinder testing, as the duplicates tend to be repeatedly triggered. To tackle this, we attempted to disable the construction of bug-inducing elements, and also ignored known discrepancy patterns before the reported bug was resolved. 

\paragraph{Found bugs overview.} As shown in Table \ref{tab:bugs}, we successfully found \bugnum{} unique bugs in total, 15 in BaseX, 6 in eXist-DB, 4 in Saxon, and 2 in the commercial DBMS. As detailed subsequently, we could have reported additional bugs for eXist-DB and the commercial DBMS, but refrained from doing so due to the high number of unfixed bugs for eXist-DB, and lack of developer feedback for the commercial DBMS.
The bug-inducing test cases we found were not covered by the W3C qt3 test suite~\cite{W3Cqt3}, which contains around 30,000 tests for XPath and XQuery---Saxon 11.1 passes all applicable tests in the W3C qt3 test suite~\cite{Saxon-31-conformance}. Out of the \bugnum{} bugs found, the majority, \logicbugnum{} bugs, were logic bugs. Based on developer feedback, we learned that among the \fixedbugnum{} fixed bugs, at least \optimizationbugnum{} bugs were due to incorrect optimizations. We detected the remaining bugs through unexpected errors. All systems we tested were implemented in Java, so we did not observe any crash bugs.
We did not find any bugs in PostgreSQL and libXML2, both of which are known to be robust systems.
For example, previous bug-finding efforts on testing DBMSs using SQL queries also found no logic bug in PostgreSQL~\cite{norec,tlp}.

\begin{table}[bt]
  \caption{Bugs found by XPress}
  \label{tab:bugs}
  \begin{tabular}{lrrrr}
    \toprule
    XML Processor&Fixed&Confirmed&Reported&Total\\
    \midrule
    BaseX & 15 & 0 & 0 & 15\\
    eXist-DB & 1 & 5 & 0 & 6 \\
    Saxon & 4 & 0 & 0 & 4 \\
    Commercial DBMS & 0 & 0 & 2 & 2 \\
  \bottomrule
\end{tabular}
\end{table}

\begin{table}[bt]
  \caption{Category of Bugs found by XPress}
  \label{tab:category}
  \begin{tabular}{lrrrr}
    \toprule
    XML Processor&Logic Bugs&Internal Errors\\
    \midrule
    BaseX & 10 & 5\\
    eXist-DB & 5 & 1\\
    Saxon & 2 & 2\\
    Commercial DBMS & 2 & 0\\
  \bottomrule
\end{tabular}
\end{table}

\paragraph{Small-scope hypothesis}
We observed that the reported bugs are mainly reproducible by short test cases. 70\% of all the reported cases can be reproduced with an XML document that consists of only one node and 91\% of XPath expression consists of only one section. The average length of the XML documents in the reported test cases was 12 characters and XPath expressions 30 characters. This phenomenon is known as the small-scope hypothesis~\cite{andoni2003evaluating}, and this observation has been exploited in testing work that systematically generates small test inputs~\cite{mohan2018finding}.

\paragraph{Developer reception.} Developer feedback is an important indicator of the bugs' importance. A core developer of BaseX stated \emph{"Thanks for sharing the bug reports with us. I appreciate that, they’re definitely helpful."}\footnote{\url{https://www.mail-archive.com/basex-talk@mailman.uni-konstanz.de/msg15173.html}} All 15 bugs reported to BaseX were resolved within one month---10 bugs were resolved even within 24 hours. This indicates not only that the team was fast in resolving bugs, but also that the bug reports were considered valuable. Due to the timely fixes of the BaseX team, we invested most time and effort in testing BaseX. After encouragement from the developers of BaseX, we contributed the bug-inducing test cases to the W3C XQuery and XPath test suite~\cite{W3Cqt3}. Most bugs submitted to eXist-DB have not yet been fixed, which is likely the result of the many open issues (over 400). Nevertheless, the developers from eXist-DB confirmed the bugs quickly and also expressed appreciation towards the bug reports \emph{"thank you for finding and reporting."}\footnote{\url{https://www.mail-archive.com/basex-talk@mailman.uni-konstanz.de/msg15204.html}}\footnote{\url{https://github.com/eXist-db/exist/issues/4830}} Because the reported bugs remained unfixed for over two months, we stopped testing and reporting to eXist-DB after reporting the first few found inconsistencies due to the difficulties of filtering out duplicate bugs.
We believe that XPress has the ability to find more bugs in eXist-DB after the known bugs are resolved. 
Similarly, for the commercial DBMS, since the developers did not follow up on the bugs that we reported, we stopped testing this DBMS.
For Saxon, all four bugs reported were resolved quickly within one week's time. 

\paragraph{Selected bugs} Below, we give a few selected examples of bugs found by XPress to illustrate its bug-finding capability.

\begin{figure}[bt]
  \centering
  \includegraphics[width=\linewidth]{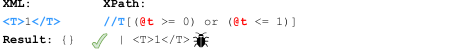}
  \caption{Incorrect optimization of comparison conditions.}
  \label{fig:bug1}
\end{figure}

\paragraph{Incorrect optimization of comparison conditions.} Figure \ref{fig:bug1} shows a fixed bug that we reported to BaseX.\footnote{\url{https://github.com/BaseXdb/basex/issues/2190}} The XPath expression selects all \hli{T} nodes with attribute \hli{@t} that satisfies \hli{@t >= 0 or @t <= 1}. When \hli{@t} exists and is a numeric value, this is a condition that always evaluates to true. Therefore, an optimization in BaseX rewrote the predicate to true. However, when \hli{@t} does not exist for node \hli{T}, \hli{@t} evaluates to an empty sequence and returns false for both \hli{@t >= 0} and \hli{@t <= 1}. Before we reported this bug, this case was overlooked and resulted in an incorrect optimization.

\begin{figure}[bt]
  \centering
  \includegraphics[width=\linewidth]{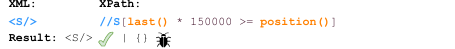}
  \caption{Arithmetic overflow in pre-check conditions.}
  \label{fig:bug2}
\end{figure}

\paragraph{Arithmetic overflow in pre-check conditions.} Figure \ref{fig:bug2} shows a fixed bug that we reported to BaseX.\footnote{\url{https://github.com/BaseXdb/basex/issues/2220}} \hli{last()} and \hli{position()} returns the context size and the context position from the dynamic context respectively. In the context XML document, the prefix \hli{//S} selects only one node, and therefore both \hli{last()} and \hli{position()} return \hli{1}. Therefore, the condition is true and node \hli{S} should be selected. In BaseX, an empty result set was returned. The problem was related to optimization for positional arguments in conditional comparisons. BaseX substituted \hli{last()} with the greatest theoretical \hli{last()} value and checked if the condition could evaluate to true. If not, the condition could not be satisfied regardless of the actual context and could be rewritten to false to reduce context analysis. When calculating the multiplication, as the theoretical maximum value for \hli{last()} is a big integer, calculating the expression with \hli{long} instead of \hli{double} caused an overflow and 
produced the incorrect result. 

\begin{figure}[bt]
  \centering
  \includegraphics[width=\linewidth]{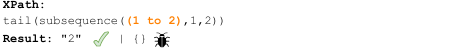}
  \caption{Result of tail after subsequence off by one.}
  \label{fig:bug3}
\end{figure}

\paragraph{Result of tail after subsequence off by one.} Figure \ref{fig:bug3} shows a fixed bug that we reported to eXist-DB.\footnote{\url{https://github.com/eXist-db/exist/issues/4830}} \hli{1 to 2} creates an integer sequence consisting of 1 and 2. The \hli{subsequence()} function in this example selects two elements starting from index 1, and the \hli{tail()} function returns a new sequence excluding the first element of the input sequence. The correct result is to return 2. Unexpectedly, eXist-DB returned an empty result set. This was caused by a mistake when processing a call to \hli{tail} that has a call to \hli{subsequence} as an argument, which incorrectly reduced the ending index by 1. 

\begin{figure}[bt]
  \centering
  \includegraphics[width=\linewidth]{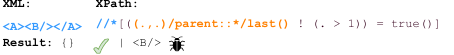}
  \caption{Incorrect reduce in positional expressions.}
  \label{fig:bug4}
\end{figure}

\paragraph{Incorrect reduce in positional expressions.} Figure \ref{fig:bug4} shows a fixed bug that we reported to Saxon.\footnote{\url{https://saxonica.plan.io/issues/6093?pn=1\#change-24136}} The dot (\hli{.}) stands for the current context in XPath expressions. For node B, \hli{(., .)/parent::*} selects the single node \hli{A} as the parent. Therefore, \hli{last() = 1} and the condition evaluates to false. Saxon unexpectedly returned the node \hli{B}. The \hli{=} operator is considered to be an unordered operator, which does not require operands to be sorted. In Saxon, an optimization was applied to eschew removing duplicate nodes when evaluating the sub-expression, which resulted in \hli{A} being selected twice and \hli{last()} evaluated to 2. After we found and reported the bug, a patch was applied by the developers to remove the duplicates, when the left operand of \hli{=} is positional sensitive.



\subsection{Efficiency} \label{efficiency}
\rnew{\paragraph{Existing-generator baselines. } We considered the only two---to the best of our knowledge---approaches to generate XPath expressions. Neither of them was specifically designed to be combined with a XPath test oracle. XQgen \cite{Wu09} generates XPath queries for micro-benchmarking. Its generated predicates only check for sub-element existence. The XQuery generator designed by Todic and Uzelac \cite{Todic12} generates XPath queries for automatically testing index support in DBMSs. Given that indexes apply only to sargable queries (\emph{i.e.}, simple comparisons), the expressions it generates are simple. Both approaches generate XPath expressions based on an XML schema, while XPress generates XPath expressions based on the actual XML document. Based on this, we expect both of them to have low applicability for our differential-testing approach. Given that neither implementations are publicly available, we re-implemented them based on the description in the papers.}

\rnew{\paragraph{Self-constructed baselines. } We also constructed our own baselines to investigate the efficiency of the separate components of XPress.}\excludecomment{ Due to the lack of publicly available automated testing baselines, we constructed our own baselines to investigate the efficiency of the separate components of XPress. Our major contributions to XPath generation are}\rnew{ XPress has two main components, namely} (1) the targeted predicate generation by using the targeted node to refer to existing nodes and attributes and (2) the predicate rectification to avoid empty result sets. To evaluate the effect of the components individually, we enabled them individually to test whether they improve XPress's bug detection efficiency. 

\paragraph{Configurations.} We considered four configurations\excludecomment{ in our efficiency evaluation.}\rnew{for our self-constructed baselines.} Apart from our proposed approach introduced in Section \ref{xpath} as (1) \emph{Targeted}, we derive configuration (2)  \emph{Targeted without Rectification}, (3)  \emph{Untargeted with Rectification}, and (4)  \emph{Untargeted without Rectification}. In (2)  \emph{Targeted without Rectification}, we disable the rectification process, which would otherwise ensure targeted node selection. Since selecting a targeted node for predicate generation guidance always requires at least one node in the result set, we stop generating new sections after an empty result set is produced. In (3)  \emph{Untargeted with Rectification}, we generate predicates without using targeted node information to supply parameters that reference existent context and trigger corner cases for function nodes, while keeping the rectification to ensure that at least one node from the candidate set is included in the result set. In (4)  \emph{Untargeted without rectification}, we remove both components to generate predicates randomly, while omitting rectification.

\paragraph{Methodology.}
We set each\excludecomment{ configuration}\rnew{baseline} to run for 24 hours~\cite{Klees18}. We repeated each experiment 10 times to account for potential performance deviations, and report the arithmetic mean for all metrics.
As our testing target, we selected BaseX 10.4, which is the BaseX version that we first started testing.
The reason for selecting BaseX as a representative is that we found most bugs in BaseX and all bugs were fixed, allowing us to determine the number of \emph{unique} bugs we found in a testing campaign by deduplicating bug-inducing test cases automatically.
Specifically, given two bug-inducing test cases, we could determine whether they trigger the same underlying bug by identifying their fix commits; only if their associated fix commit are different, do we consider the bugs unique.
This is a best-effort technique, as, for example, one fix commit might address multiple bugs.
We disabled the generation of the \hli{has-children} functions as well as using relative XPath expressions in predicates, as they consistently lead to crashes, triggering known bugs.

\begin{figure}[bt]
  \centering
  \includegraphics[width=0.85\linewidth]{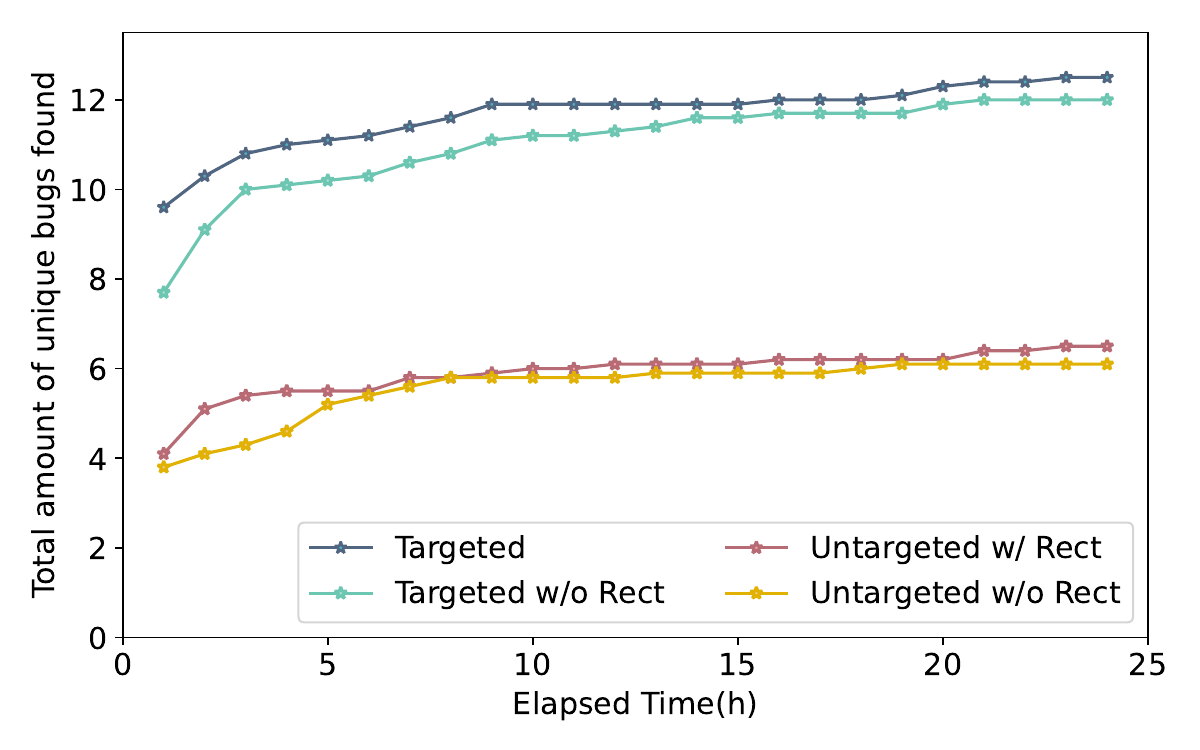}
  \caption{Average number of unique bugs found under different configurations in 24 hours across 10 runs.}
  \label{fig:bug collection}
\end{figure}

\begin{table}[bt]
\small
  \caption{Average bug report collection under different configurations in 24 hours across 10 runs.}
  \label{tab:diag}
  \begin{tabular}{p{2.3cm}R{1.0cm}R{1.3cm}R{0.85cm}R{1.33cm}}
    \toprule
    Config&Total cases&Differences detected&Unique bugs&Non-empty result\\
    \midrule
    Targeted & 6.6M & 11.8K & 12.5 & 100\%\\
    Targeted w/o Rect & 9.4M & 10.2K & 12 & 66\%\\
    Untargeted w/ Rect & 8.8M & 1.4K & 6.5 & 100\%\\
    Untargeted w/o Rect & 13.5M & 0.6K & 6.1 & 44\%\\
  \bottomrule
\end{tabular}
\end{table}


\rnew{\paragraph{Results of existing generators.} Neither XQGen nor the Combined XML/XQuery generator found bugs in our experiment. This is expected, as previously proposed approaches were not designed for automated testing. As mentioned above, XQGen generates predicates that only check for element existence. The XQuery generator designed by Todic and Uzelac generates simple predicates that include at most one comparison operator. }

\paragraph{Results \rnew{of different configurations.}} As Figure~\ref{fig:bug collection} shows, our proposed approach, \emph{Targeted} outperforms the other configurations. Within 24 hours, it found the most number of unique bugs (namely 12.5). 
Both configurations with targeted generation clearly outperformed the untargeted approaches, while rectification shows\excludecomment{ only slight improvements} \rnew{a similar performance} in the speed of bug detection. As shown in Table \ref{tab:diag}, both targeted generation and rectification reduce the testing throughput, as they obtain intermediate results using the XML processor under test. 
Despite generating only 50\% of the number of test cases as compared to (4) \emph{Untargeted without Rectification}, (1) \emph{Targeted} detected 20$\times$ more bug-inducing test cases and 2$\times$ more unique bugs. The results show that selecting a target node to guide the XPath generation process improves testing efficiency significantly\rnew{.}\excludecomment{, while rectification also helps slightly in both targeted and untargeted cases.} \rnew{As observed above when discussing the small-scope hypothesis, most of the bugs that we found can be reproduced using a single section, explaining the limited effectiveness of rectification. However, we still believe that rectification is an important component, since without it, bugs requiring multiple sections with non-empty results could hardly be found.}

\paragraph{Code coverage.} We collected code coverage for three processors' core modules for XPress for 24 hours~\cite{Klees18} of execution. The result is shown in Table \ref{tab:coverage}. To put the numbers in relation, we collected coverage also for the projects' test suites; Saxon has no publicly available test suites and is therefore excluded. For the three XML processors, the line coverage ranged from 15\% to 20\%, and the branch coverage ranged from 10\% to 16\%. The coverage percentages are low, which is expected. The main reason for low code coverage is that XML processors typically also have other components than XPath processing.
Taking BaseX as an example, around 21\% of uncovered code was GUI-related,  10\% was due to lack of full-text functionality support, and 5\% were database commands. In Saxon, as another example, XSLT modules have not been covered.
A further 18\% uncovered code in BaseX involved unimplemented functions; it would be straightforward to implement many additional ones, such as math functions, but the many functions available would make this a tedious task.
In Section \ref{historical}, we detail unsupported XPath features, implementing which might allow us to find more bugs. XPress's test-case generation process primarily aims at generating semantically valid expressions, which results in low error-checking branch coverage, quantifying which is difficult, as the relevant code is spread throughout the code base.

\begin{table}[bt]
  \caption{Code coverage of tested systems in 24 Hours}
  \label{tab:coverage}
  \begin{tabular}{rrrrrrr}
    \toprule
    \multirow{2}{*}{Approach} & \multicolumn{2}{c}{BaseX} & \multicolumn{2}{c}{eXist} & \multicolumn{2}{c}{Saxon}\\
    \cmidrule(lr){2-3}\cmidrule(lr){4-5}\cmidrule(lr){6-7}
    \: & Line & Branch & Line & Branch & Line & Branch \\
    XPress & 20\% & 16\% & 18\% & 10\% & 15\% & 10\%\\
    Unit Tests & 67\% & 58\% & 52\% & 47\% & - & -\\
  \bottomrule
\end{tabular}
\end{table}

\subsection{Comparison to the State of the Art} \label{comparison}

We are aware of only one automated testing approach that has been proposed to test XML processors~\cite{Todic12}. It tackled the test oracle problem by using differential testing by comparing the results of Microsoft's SQLServer with and without using indexes. Their approach was specifically designed to test SQLServer's index support and is not publicly available. Due to the narrow testing scope, and since the tool is not publicly available, we could not conduct experiments to directly compare the approaches. However, we further extended our tool to support differential testing with index configurations. Both approaches are complementary, as XPress could not only use differential testing among various XML processors, but also create or omit indexes to find additional bugs. 

\paragraph{Index support in BaseX, eXist-DB, Saxon, and libxml2} Database indexes are data structures built to speed up data retrieval~\cite{li2001indexing} and are DBMS-specific. Not all XML processors are DBMSs---as in-memory processors, Saxon and libxml2 lack support for indexes. BaseX and eXist-DB both enable structural indexes, such as storing all distinct paths of nodes by default. For value indexes to optimize querying on content values, BaseX creates text index and attribute index automatically. Users can further define additional indexes. Additionally, BaseX provides token indexes, which apply to specific functions, such as \hli{contains-token}. 
eXist supports range indexes, which could be defined for specific nodes or attributes to speed up related comparison searches on their contents. 

\paragraph{Methodology} We tested eXist's range index and BaseX's token index using the XPath expression generation approach as described in Section \ref{xpath}. Due to the found unfixed bugs in eXist, we conducted differential testing within eXist by checking the results with and without range index definition. For BaseX, we defined a token index and compared its results directly with the results of Saxon. 

\begin{figure}[bt]
  \centering
  \includegraphics[width=\linewidth]{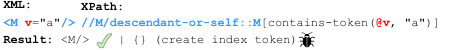}
  \caption{Found bug with token index in BaseX.}
  \label{token bug}
\end{figure}

\paragraph{Results.} Throughout the testing method, we detected one additional bug for BaseX\footnote{\url{https://github.com/BaseXdb/basex/issues/2222}} and found no additional bugs in eXist.  We reported the found bug shown in Figure \ref{token bug} to the BaseX developers, who quickly fixed it. The query selects all nodes with tag name \hli{M} in the document which holds attribute \hli{v} that contains token \hli{"a"}. BaseX returned node \hli{M} without token index, as expected, while unexpectedly returning an empty result set when not using an index.
Overall, while the results suggest that using or removing indexes might find additional bugs, doing so had low effectiveness.
A potential explanation could be that our test-case generation approach does not consider when indexes could be applied, which might result in low testing efficiency.

\subsection{Analysis of BaseX Historical Bug Reports} \label{historical}
Unlike formal verification approaches, automatic testing approaches might miss bugs in the system tested. Due to the lack of ground truth, we cannot generally determine which bugs are overlooked by our approach. However, as a best-effort approach, we studied historical bug reports in order to determine whether XPress could have found them. 

\paragraph{Bug reports}
We analyzed all historical BaseX bug reports in its GitHub bug tracker. We selected BaseX, because the majority of issues are closed (1618 out of 1640). The issue tracker of BaseX is used for confirmed bug reports filtered from reports from the mailing list, and the BaseX maintainers carefully label and document them. For these reasons, it was easy to identify and classify the underlying problem of each bug report. 

\paragraph{Methodology.} We manually analyzed all historical bug issues until 2023 Apr 17 in BaseX, which were 1597 issues, after excluding the issues we reported. To confine the study of bug reports within the scope of XPath, we selected bug reports triggered by only XPath expressions. To determine whether a bug could be theoretically found by XPress, we mainly checked three aspects of the reports. For XPress to cover the test case, both the XML document and the XPath expression in the test case should not include any unimplemented functions or language features. Second, we could construct the sections and the predicate tree structure of XPress for involved predicates to form the pattern of the bug-inducing XPath expression. Third, XML processors should disagree on the result set. Note that this is a best-effort approach, because we might both incorrectly conclude that XPress might find a bug (\emph{e.g.}, it might be unlikely that the test case would be generated in practice) or incorrectly conclude that a bug cannot be found even when a different test-case within the reach of XPress would trigger the same underlying bug.

\paragraph{Results.} Out of the total 78 bugs that we collected, we identified 20 bugs that could have been detected by XPress. For the other 58 bugs, we identified 4 kinds of bugs that XPress would have failed to find, namely due to (1) unimplemented functionalities (51 cases), (2) invalid inputs where the expected result would be an error (6 cases), (3) processors producing different results (2 cases), and (4) miscellaneous other issues (6 cases). Bugs belonging to more than one group are included in all involved groups. The differential testing oracle fails to detect the bugs with processors producing different results, while we consider the other categories mostly as implementation limitations in test-case generation. Therefore, out of all 78 bugs, 76 bugs (97\%) could be detected through differential testing. This further demonstrates the effectiveness of employing a differential testing oracle for XPath-related testing.

\paragraph{Unimplemented functionalities} Most uncovered bug reports are due to unimplemented functionalities. Unsupported functions include constructors defined by the XML or XPath language standards, array and map functions, and also constructors of derived datatypes~\cite{xml-datatypes}, such as \hli{xs:NMtokens}. Given enough time, it would be straightforward to implement them in XPress. 
For/while loops, variable declaration, if-else conditional expressions, and self-defined functions are also unimplemented.
These could be supported based on approaches that have been proposed in the context of compiler testing~\cite{yang2011finding, livinskii2023fuzzing}.
Neither the XML documents nor XPath expressions that XPress  constructs involve namespaces, which allow distinguishing items with the same tag name. They could be integrated into the XPress test-case generator. By implementing all these features, an additional 38 bugs (48\%) could have been found. 

\paragraph{Expected errors} Bug reports grouped into \textit{expected is error} refers to invalid test cases, which are successfully executed instead of throwing an error. XPress constructs both syntactically and semantically valid expressions and therefore could not detect bugs within this category. However, the differential testing oracle could detect these bugs by comparing the errors of the different XML processors.

\paragraph{Different results} The different result category contains queries for which different processors intentionally produce different results, which shows the limitation of the differential testing oracle. One example is the function \hli{id}, which selects nodes with \hli{xml:id} attributes. BaseX takes attributes named as \hli{id} as \hli{xml:id} attributes, while Saxon and eXist-DB require an explicit declaration.  


\section{Related Work}

While various related approaches to our work exist, to the best of our knowledge, we propose the first general approach to testing XML processors to find logic bugs. As discussed above, the most closely related work proposed testing the index support of SQLServer in the context of XPath and XQuery~\cite{Todic12}\rnew{, which, to the best of our knowledge, is the only work that has tackled the test-oracle problem for XML processors, but is limited in scope}. 

\paragraph{Testing XPath functionality} Various approaches to benchmarking XPath implementations or test suites for them have been proposed, the most representative being XPathMark and the W3C qt3 test suite. XPathMark~\cite{Franceschet05} is a benchmark for testing XML processors' XPath standard 1.0 functionality, containing both correctness as well as performance tests. 
The W3C qt3 test suite developed by the W3C XQuery and XSLT Working Groups~\cite{W3Cqt3} contains around 30,000 tests for XPath and XQuery targeting XPath 3.0 and later versions, which cover a broad range of functions and expressions. 

\paragraph{XML-related automated synthetic data generation} Previous works have proposed approaches for automatically generating XML-related data, such as XML documents, XPath, and XQuery expressions. 
Aboulnaga et al. proposed an XML document generator to generate synthetic, but complex, structured XML data by introducing recursion and repetition on tag name assignment and controlling the element frequency distribution~\cite{Abou01}. Rychnovský and Holubová proposed an approach to generate XML documents related to given XPath queries from a specific XML schema to improve query efficiency~\cite{Rychnovsky15}, which is useful for developers to create micro-benchmarks for testing performance over certain XPath expressions. XQGen~\cite{Wu09} is a tool for generating XPath queries that conform to a given XML schema, allowing users to specify multiple parameters, such as the percentage of empty queries desired and the percentage of queries with predicates. XPath generated by XQGen includes only direct node tests without introducing complex expressions, such as axes or function transformations. Similarly, the XQuery generator designed by Todic and Uzelac~\cite{Todic12} includes XQuery FLWOR expressions, but the logic predicate consists only of simple operations, such as value comparisons.  
Neither of these works tackled the test oracle problem\rnew{, and, as indicated by the results in Section~\ref{comparison}, given their different focus, they cannot be effectively combined with a differential testing oracle.}

\paragraph{Targeted test case generation} Many testing tools guide their test case generation process to improve testing efficiency, for random approaches such as random byte mutation used in fuzzing approaches generate a large proportion of invalid queries~\cite{zhong2020squirrel}. DynSQL~\cite{jiang2023dynsql} guides the fuzzing process of DBMSs towards increased code coverage and high statement validity. \textsc{APOLLO}~\cite{jung2019apollo} is a system for detecting performance regression bugs in DBMSs. It increases the probability of including components from previously encountered performance issues. Cynthia~\cite{Sotiropoulos21} was proposed to test Object Relational Mappers (ORMs) and generates targeted databases dependent on generated abstract SQL queries, which are likely to return non-empty results. Query Plan Guidance (QPG)~\cite{ba2023testing} guides testing towards exploring more unique query plans.

\rnew{
\paragraph{Pivoted Query Synthesis}
The \emph{targeted node} in XPress was inspired by the \emph{pivot row} in \emph{Pivoted Query Synthesis (PQS)}~\cite{rigger2020testing}, which was originally proposed to test relational DBMSs.
PQS' and XPress' commonality is that they select a random element, in PQS, a row in the database, while for XPress, a node in an XML document, based on which they generate a query that is guaranteed to fetch the element.
However, both the purpose and use of the targeted node and pivot row differ.
In PQS, the pivot row is used both for test-case generation and to construct the test oracle, by evaluating an expression and ensuring that it evaluates to true for the pivot row so that it can be used in a query that is guaranteed to fetch the row.
Doing so requires a naive reimplementation of all the DBMSs' operators that should be tested, which incurs a high implementation effort, as highlighted in follow-up work~\cite{10.1145/3368089.3409710   }.
In XPress, the targeted node is used only for test-case generation, to improve testing efficiency and to ensure non-empty intermediate results; to this end, XPress uses the XML processor to determine the result of the expression, rather than requiring the reimplementation of operators.
In addition, for predicate rectification, XPress provides operator-specific rules, rather than relying on a generic one, aiming to generate more interesting test cases.
The high-level idea of a pivot element also inspired other works; for example, recent work on Android testing introduced the concept of a \emph{pivot layout}~\cite{SuTing2021}.
}

\section{Conclusion}
This paper has presented a general automated testing approach for detecting XPath-related logic bugs in XML processors. We demonstrate that differential testing is applicable in this domain, since XML processors widely adhere to the XPath standards. To generate interesting XPath queries, our approach selects a so-called targeted node to guide predicate generation and predicate rectification to ensure the inclusion of that node. Our evaluation shows that this improves the number of bugs detected in 24 hours to 2$\times$ as compared to random generation. More importantly, we have successfully detected \bugnum{} previously unknown, unique bugs in six mature XML processing systems. 
We believe that this high number is surprising, given that XML processors are an essential part of our computing infrastructure, with the first XPath standard having been proposed more than 20 years ago, and the systems that we have tested having been maintained for at least 15 years.
We believe that XPress, given its simplicity and generality, has a high chance of being integrated into the toolbox of XML processor developers.
Furthermore, we believe that our work might inspire testing approaches for other XML standards, such as XQuery or XSLT.

\begin{acks}
This research was supported by a Ministry of Education (MOE) Academic Research Fund (AcRF) Tier 1 grant.
\end{acks}

\balance
\bibliographystyle{ACM-Reference-Format}
\bibliography{sample-base}


\begin{thebibliography}{47}


\ifx \showCODEN    \undefined \def \showCODEN     #1{\unskip}     \fi
\ifx \showDOI      \undefined \def \showDOI       #1{#1}\fi
\ifx \showISBNx    \undefined \def \showISBNx     #1{\unskip}     \fi
\ifx \showISBNxiii \undefined \def \showISBNxiii  #1{\unskip}     \fi
\ifx \showISSN     \undefined \def \showISSN      #1{\unskip}     \fi
\ifx \showLCCN     \undefined \def \showLCCN      #1{\unskip}     \fi
\ifx \shownote     \undefined \def \shownote      #1{#1}          \fi
\ifx \showarticletitle \undefined \def \showarticletitle #1{#1}   \fi
\ifx \showURL      \undefined \def \showURL       {\relax}        \fi
\providecommand\bibfield[2]{#2}
\providecommand\bibinfo[2]{#2}
\providecommand\natexlab[1]{#1}
\providecommand\showeprint[2][]{arXiv:#2}

\bibitem[XPa(1999)]%
        {XPath-1.0-ref}
 \bibinfo{year}{1999}\natexlab{}.
\newblock \bibinfo{booktitle}{\emph{XML Path Language (XPath) Version 1.0 W3C
  Recommendation}}.
\newblock
\urldef\tempurl%
\url{https://www.w3.org/TR/1999/REC-xpath-19991116/}
\showURL{%
Retrieved July 17, 2023 from \tempurl}


\bibitem[xml(2004)]%
        {xml-datatypes}
 \bibinfo{year}{2004}\natexlab{}.
\newblock \bibinfo{booktitle}{\emph{XML Schema Part 2: Datatypes Second Edition
  - Built-in datatypes}}.
\newblock
\urldef\tempurl%
\url{https://www.w3.org/TR/xmlschema-2/#built-in-datatypes}
\showURL{%
Retrieved July 17, 2023 from \tempurl}


\bibitem[ebn(2008)]%
        {ebnf-notation}
 \bibinfo{year}{2008}\natexlab{}.
\newblock \bibinfo{booktitle}{\emph{EBNF notation from the W3C Extensible
  Markup Language (XML) 1.0 (Fifth Edition)}}.
\newblock
\urldef\tempurl%
\url{https://www.w3.org/TR/REC-xml/}
\showURL{%
Retrieved July 17, 2023 from \tempurl}


\bibitem[XPa(2014a)]%
        {XPath-3.0-ref}
 \bibinfo{year}{2014}\natexlab{a}.
\newblock \bibinfo{booktitle}{\emph{XML Path Language (XPath) 3.0 W3C
  Recommendation}}.
\newblock
\urldef\tempurl%
\url{https://www.w3.org/TR/xpath-30/}
\showURL{%
Retrieved July 17, 2023 from \tempurl}


\bibitem[XPa(2014b)]%
        {XPath-3.0-functions-ref}
 \bibinfo{year}{2014}\natexlab{b}.
\newblock \bibinfo{booktitle}{\emph{XPath and XQuery Functions and Operators
  3.0 W3C Recommendation}}.
\newblock
\urldef\tempurl%
\url{https://www.w3.org/TR/xpath-functions-30/}
\showURL{%
Retrieved July 17, 2023 from \tempurl}


\bibitem[XQu(2017)]%
        {XQuery-ref}
 \bibinfo{year}{2017}\natexlab{}.
\newblock \bibinfo{booktitle}{\emph{XQuery 3.1: An XML Query Language W3C
  Recommendation}}.
\newblock
\urldef\tempurl%
\url{https://www.w3.org/TR/xquery-31/}
\showURL{%
Retrieved July 17, 2023 from \tempurl}


\bibitem[XSL(2017)]%
        {XSLT-ref}
 \bibinfo{year}{2017}\natexlab{}.
\newblock \bibinfo{booktitle}{\emph{XSL Transformations (XSLT) Version 3.0 W3C
  Recommendation}}.
\newblock
\urldef\tempurl%
\url{https://www.w3.org/TR/xslt-30/}
\showURL{%
Retrieved July 17, 2023 from \tempurl}


\bibitem[Bas(2023)]%
        {BaseX}
 \bibinfo{year}{2023}\natexlab{}.
\newblock \bibinfo{booktitle}{\emph{BaseX}}.
\newblock
\urldef\tempurl%
\url{https://basex.org/}
\showURL{%
Retrieved July 31, 2023 from \tempurl}


\bibitem[DBE(2023)]%
        {DBEngines}
 \bibinfo{year}{2023}\natexlab{}.
\newblock \bibinfo{booktitle}{\emph{DB-Engines Ranking}}.
\newblock
\urldef\tempurl%
\url{https://db-engines.com/en/ranking}
\showURL{%
Retrieved July 6, 2023 from \tempurl}


\bibitem[eXi(2023a)]%
        {eXist-DB}
 \bibinfo{year}{2023}\natexlab{a}.
\newblock \bibinfo{booktitle}{\emph{eXist-DB}}.
\newblock
\urldef\tempurl%
\url{http://exist-db.org/exist/apps/homepage/index.html}
\showURL{%
Retrieved July 31, 2023 from \tempurl}


\bibitem[eXi(2023b)]%
        {eXist-ref}
 \bibinfo{year}{2023}\natexlab{b}.
\newblock \bibinfo{booktitle}{\emph{eXist DB reference page}}.
\newblock
\urldef\tempurl%
\url{http://exist-db.org/exist/apps/homepage/references.html}
\showURL{%
Retrieved July 6, 2023 from \tempurl}


\bibitem[lib(2023)]%
        {libXML2}
 \bibinfo{year}{2023}\natexlab{}.
\newblock \bibinfo{booktitle}{\emph{libXML2}}.
\newblock
\urldef\tempurl%
\url{https://gitlab.gnome.org/GNOME/libxml2}
\showURL{%
Retrieved July 31, 2023 from \tempurl}


\bibitem[MyS(2023)]%
        {MySQL}
 \bibinfo{year}{2023}\natexlab{}.
\newblock \bibinfo{booktitle}{\emph{MySQL}}.
\newblock
\urldef\tempurl%
\url{https://www.mysql.com/}
\showURL{%
Retrieved July 31, 2023 from \tempurl}


\bibitem[Ora(2023)]%
        {Oracle}
 \bibinfo{year}{2023}\natexlab{}.
\newblock \bibinfo{booktitle}{\emph{Oracle Database}}.
\newblock
\urldef\tempurl%
\url{https://www.oracle.com/database/}
\showURL{%
Retrieved July 31, 2023 from \tempurl}


\bibitem[Pos(2023)]%
        {PostgreSQL}
 \bibinfo{year}{2023}\natexlab{}.
\newblock \bibinfo{booktitle}{\emph{PostgreSQL}}.
\newblock
\urldef\tempurl%
\url{https://www.postgresql.org/}
\showURL{%
Retrieved July 31, 2023 from \tempurl}


\bibitem[Sax(2023a)]%
        {Saxon-ref}
 \bibinfo{year}{2023}\natexlab{a}.
\newblock \bibinfo{booktitle}{\emph{Saxon home page}}.
\newblock
\urldef\tempurl%
\url{https://saxonica.com/html/welcome/welcome.html}
\showURL{%
Retrieved July 6, 2023 from \tempurl}


\bibitem[Sax(2023b)]%
        {Saxon-31-conformance}
 \bibinfo{year}{2023}\natexlab{b}.
\newblock \bibinfo{booktitle}{\emph{Saxon XQuery 3.1 conformance page}}.
\newblock
\urldef\tempurl%
\url{https://www.saxonica.com/documentation12/#!conformance/xquery31}
\showURL{%
Retrieved July 13, 2023 from \tempurl}


\bibitem[Sax(2023c)]%
        {Saxon}
 \bibinfo{year}{2023}\natexlab{c}.
\newblock \bibinfo{booktitle}{\emph{Saxonica}}.
\newblock
\urldef\tempurl%
\url{https://saxonica.com/}
\showURL{%
Retrieved July 31, 2023 from \tempurl}


\bibitem[W3C(2023)]%
        {W3Cqt3}
 \bibinfo{year}{2023}\natexlab{}.
\newblock \bibinfo{booktitle}{\emph{W3C qt3 test suite github repository}}.
\newblock
\urldef\tempurl%
\url{https://github.com/w3c/qt3tests}
\showURL{%
Retrieved July 11, 2023 from \tempurl}


\bibitem[Aboulnaga and Zhang.(2001)]%
        {Abou01}
\bibfield{author}{\bibinfo{person}{Jeffrey F.~Naughton Aboulnaga, Ashraf} {and}
  \bibinfo{person}{Chun Zhang.}} \bibinfo{year}{2001}\natexlab{}.
\newblock \showarticletitle{Generating Synthetic Complex-Structured XML Data.}
\newblock \bibinfo{journal}{\emph{WebDB.}}  \bibinfo{volume}{1}
  (\bibinfo{year}{2001}), \bibinfo{pages}{79--84}.
\newblock


\bibitem[Andoni et~al\mbox{.}(2003)]%
        {andoni2003evaluating}
\bibfield{author}{\bibinfo{person}{Alexandr Andoni}, \bibinfo{person}{Dumitru
  Daniliuc}, \bibinfo{person}{Sarfraz Khurshid}, {and} \bibinfo{person}{Darko
  Marinov}.} \bibinfo{year}{2003}\natexlab{}.
\newblock \showarticletitle{Evaluating the “small scope hypothesis”}.
\newblock


\bibitem[Ba and Rigger(2023)]%
        {ba2023testing}
\bibfield{author}{\bibinfo{person}{Jinsheng Ba} {and} \bibinfo{person}{Manuel
  Rigger}.} \bibinfo{year}{2023}\natexlab{}.
\newblock \showarticletitle{Testing Database Engines via Query Plan Guidance}.
  In \bibinfo{booktitle}{\emph{2023 IEEE/ACM 45th International Conference on
  Software Engineering (ICSE)}}. \bibinfo{pages}{2060--2071}.
\newblock
\urldef\tempurl%
\url{https://doi.org/10.1109/ICSE48619.2023.00174}
\showDOI{\tempurl}


\bibitem[Chen et~al\mbox{.}(2016)]%
        {Chen16}
\bibfield{author}{\bibinfo{person}{Yuting Chen}, \bibinfo{person}{Ting Su},
  \bibinfo{person}{Chengnian Sun}, \bibinfo{person}{Zhendong Su}, {and}
  \bibinfo{person}{Jianjun Zhao}.} \bibinfo{year}{2016}\natexlab{}.
\newblock \showarticletitle{Coverage-Directed Differential Testing of JVM
  Implementations}. In \bibinfo{booktitle}{\emph{Proceedings of the 37th ACM
  SIGPLAN Conference on Programming Language Design and Implementation}} (Santa
  Barbara, CA, USA) \emph{(\bibinfo{series}{PLDI '16})}.
  \bibinfo{publisher}{Association for Computing Machinery},
  \bibinfo{address}{New York, NY, USA}, \bibinfo{pages}{85–99}.
\newblock
\showISBNx{9781450342612}
\urldef\tempurl%
\url{https://doi.org/10.1145/2908080.2908095}
\showDOI{\tempurl}


\bibitem[Dyck(2002)]%
        {Dewitt}
\bibfield{author}{\bibinfo{person}{Timothy Dyck}.}
  \bibinfo{year}{2002}\natexlab{}.
\newblock \bibinfo{booktitle}{\emph{DB Test Pioneer Makes History}}.
\newblock
\urldef\tempurl%
\url{https://www.eweek.com/development/db-test-pioneer-makes-history/}
\showURL{%
Retrieved July 31, 2023 from \tempurl}


\bibitem[Franceschet(2005)]%
        {Franceschet05}
\bibfield{author}{\bibinfo{person}{Massimo Franceschet}.}
  \bibinfo{year}{2005}\natexlab{}.
\newblock \showarticletitle{XPathMark: An XPath Benchmark for the XMark
  Generated Data}. In \bibinfo{booktitle}{\emph{Database and XML
  Technologies}}, \bibfield{editor}{\bibinfo{person}{St{\'e}phane Bressan},
  \bibinfo{person}{Stefano Ceri}, \bibinfo{person}{Ela Hunt},
  \bibinfo{person}{Zachary~G. Ives}, \bibinfo{person}{Zohra Bellahs{\`e}ne},
  \bibinfo{person}{Michael Rys}, {and} \bibinfo{person}{Rainer Unland}} (Eds.).
  \bibinfo{publisher}{Springer Berlin Heidelberg}, \bibinfo{address}{Berlin,
  Heidelberg}, \bibinfo{pages}{129--143}.
\newblock
\showISBNx{978-3-540-31968-9}


\bibitem[Hua et~al\mbox{.}(2023)]%
        {hua2023gdsmith}
\bibfield{author}{\bibinfo{person}{Ziyue Hua}, \bibinfo{person}{Wei Lin},
  \bibinfo{person}{Luyao Ren}, \bibinfo{person}{Zongyang Li},
  \bibinfo{person}{Lu Zhang}, \bibinfo{person}{Wenpin Jiao}, {and}
  \bibinfo{person}{Tao Xie}.} \bibinfo{year}{2023}\natexlab{}.
\newblock \showarticletitle{GDsmith: Detecting Bugs in Cypher Graph Database
  Engines}. \bibinfo{publisher}{Association for Computing Machinery},
  \bibinfo{address}{New York, NY, USA}.
\newblock
\showISBNx{9798400702211}
\urldef\tempurl%
\url{https://doi.org/10.1145/3597926.3598046}
\showURL{%
\tempurl}


\bibitem[Jiang et~al\mbox{.}(2023)]%
        {jiang2023dynsql}
\bibfield{author}{\bibinfo{person}{Zu-Ming Jiang}, \bibinfo{person}{Jia-Ju
  Bai}, {and} \bibinfo{person}{Zhendong Su}.} \bibinfo{year}{2023}\natexlab{}.
\newblock \showarticletitle{DynSQL: Stateful Fuzzing for Database Management
  Systems with Complex and Valid SQL Query Generation}. In
  \bibinfo{booktitle}{\emph{Proceedings of the 32nd USENIX Conference on
  Security Symposium}} (Anaheim, CA, USA) \emph{(\bibinfo{series}{SEC '23})}.
  \bibinfo{publisher}{USENIX Association}, \bibinfo{address}{USA}, Article
  \bibinfo{articleno}{277}, \bibinfo{numpages}{17}~pages.
\newblock
\showISBNx{978-1-939133-37-3}


\bibitem[Jung et~al\mbox{.}(2019)]%
        {jung2019apollo}
\bibfield{author}{\bibinfo{person}{Jinho Jung}, \bibinfo{person}{Hong Hu},
  \bibinfo{person}{Joy Arulraj}, \bibinfo{person}{Taesoo Kim}, {and}
  \bibinfo{person}{Woonhak Kang}.} \bibinfo{year}{2019}\natexlab{}.
\newblock \showarticletitle{APOLLO: Automatic Detection and Diagnosis of
  Performance Regressions in Database Systems}.
\newblock \bibinfo{journal}{\emph{Proc. VLDB Endow.}} \bibinfo{volume}{13},
  \bibinfo{number}{1} (\bibinfo{date}{sep} \bibinfo{year}{2019}),
  \bibinfo{pages}{57–70}.
\newblock
\showISSN{2150-8097}
\urldef\tempurl%
\url{https://doi.org/10.14778/3357377.3357382}
\showDOI{\tempurl}


\bibitem[Kamm et~al\mbox{.}(2023)]%
        {kamm2023testing}
\bibfield{author}{\bibinfo{person}{Matteo Kamm}, \bibinfo{person}{Manuel
  Rigger}, \bibinfo{person}{Chengyu Zhang}, {and} \bibinfo{person}{Zhendong
  Su}.} \bibinfo{year}{2023}\natexlab{}.
\newblock \showarticletitle{Testing Graph Database Engines via Query
  Partitioning}. \bibinfo{publisher}{Association for Computing Machinery},
  \bibinfo{address}{New York, NY, USA}.
\newblock
\showISBNx{9798400702211}
\urldef\tempurl%
\url{https://doi.org/10.1145/3597926.3598044}
\showURL{%
\tempurl}


\bibitem[Klees et~al\mbox{.}(2018)]%
        {Klees18}
\bibfield{author}{\bibinfo{person}{George Klees}, \bibinfo{person}{Andrew
  Ruef}, \bibinfo{person}{Benji Cooper}, \bibinfo{person}{Shiyi Wei}, {and}
  \bibinfo{person}{Michael Hicks}.} \bibinfo{year}{2018}\natexlab{}.
\newblock \showarticletitle{Evaluating Fuzz Testing}.
\newblock \bibinfo{journal}{\emph{Proceedings of the 2018 ACM SIGSAC conference
  on computer and communications security}} (\bibinfo{year}{2018}).
\newblock
\urldef\tempurl%
\url{https://doi.org/10.1145/3243734.3243804}
\showDOI{\tempurl}


\bibitem[Li and Moon(2001)]%
        {li2001indexing}
\bibfield{author}{\bibinfo{person}{Quanzhong Li} {and} \bibinfo{person}{Bongki
  Moon}.} \bibinfo{year}{2001}\natexlab{}.
\newblock \showarticletitle{Indexing and Querying XML Data for Regular Path
  Expressions}. In \bibinfo{booktitle}{\emph{Proceedings of the 27th
  International Conference on Very Large Data Bases}}
  \emph{(\bibinfo{series}{VLDB '01})}. \bibinfo{publisher}{Morgan Kaufmann
  Publishers Inc.}, \bibinfo{address}{San Francisco, CA, USA},
  \bibinfo{pages}{361–370}.
\newblock
\showISBNx{1558608044}


\bibitem[Livinskii et~al\mbox{.}(2023)]%
        {livinskii2023fuzzing}
\bibfield{author}{\bibinfo{person}{Vsevolod Livinskii}, \bibinfo{person}{Dmitry
  Babokin}, {and} \bibinfo{person}{John Regehr}.}
  \bibinfo{year}{2023}\natexlab{}.
\newblock \showarticletitle{Fuzzing Loop Optimizations in Compilers for C++ and
  Data-Parallel Languages}.
\newblock \bibinfo{journal}{\emph{Proc. ACM Program. Lang.}}
  \bibinfo{volume}{7}, \bibinfo{number}{PLDI}, Article \bibinfo{articleno}{181}
  (\bibinfo{date}{jun} \bibinfo{year}{2023}), \bibinfo{numpages}{22}~pages.
\newblock
\urldef\tempurl%
\url{https://doi.org/10.1145/3591295}
\showDOI{\tempurl}


\bibitem[Mohan et~al\mbox{.}(2018)]%
        {mohan2018finding}
\bibfield{author}{\bibinfo{person}{Jayashree Mohan}, \bibinfo{person}{Ashlie
  Martinez}, \bibinfo{person}{Soujanya Ponnapalli}, \bibinfo{person}{Pandian
  Raju}, {and} \bibinfo{person}{Vijay Chidambaram}.}
  \bibinfo{year}{2018}\natexlab{}.
\newblock \showarticletitle{Finding Crash-Consistency Bugs with Bounded
  Black-Box Crash Testing}. In \bibinfo{booktitle}{\emph{13th USENIX Symposium
  on Operating Systems Design and Implementation (OSDI 18)}}.
  \bibinfo{pages}{33--50}.
\newblock


\bibitem[Rigger and Su(2020a)]%
        {norec}
\bibfield{author}{\bibinfo{person}{Manuel Rigger} {and}
  \bibinfo{person}{Zhendong Su}.} \bibinfo{year}{2020}\natexlab{a}.
\newblock \showarticletitle{Detecting Optimization Bugs in Database Engines via
  Non-Optimizing Reference Engine Construction}. In
  \bibinfo{booktitle}{\emph{Proceedings of the 28th ACM Joint Meeting on
  European Software Engineering Conference and Symposium on the Foundations of
  Software Engineering}} (Virtual Event, USA) \emph{(\bibinfo{series}{ESEC/FSE
  2020})}. \bibinfo{publisher}{Association for Computing Machinery},
  \bibinfo{address}{New York, NY, USA}, \bibinfo{pages}{1140–1152}.
\newblock
\showISBNx{9781450370431}
\urldef\tempurl%
\url{https://doi.org/10.1145/3368089.3409710}
\showDOI{\tempurl}


\bibitem[Rigger and Su(2020b)]%
        {tlp}
\bibfield{author}{\bibinfo{person}{Manuel Rigger} {and}
  \bibinfo{person}{Zhendong Su}.} \bibinfo{year}{2020}\natexlab{b}.
\newblock \showarticletitle{Finding Bugs in Database Systems via Query
  Partitioning}.
\newblock \bibinfo{journal}{\emph{Proc. ACM Program. Lang.}}
  \bibinfo{volume}{4}, \bibinfo{number}{OOPSLA}, Article
  \bibinfo{articleno}{211} (\bibinfo{date}{nov} \bibinfo{year}{2020}),
  \bibinfo{numpages}{30}~pages.
\newblock
\urldef\tempurl%
\url{https://doi.org/10.1145/3428279}
\showDOI{\tempurl}


\bibitem[Rigger and Su(2020c)]%
        {rigger2020testing}
\bibfield{author}{\bibinfo{person}{Manuel Rigger} {and}
  \bibinfo{person}{Zhendong Su}.} \bibinfo{year}{2020}\natexlab{c}.
\newblock \showarticletitle{Testing Database Engines via Pivoted Query
  Synthesis}. In \bibinfo{booktitle}{\emph{Proceedings of the 14th USENIX
  Conference on Operating Systems Design and Implementation}}
  \emph{(\bibinfo{series}{OSDI'20})}. \bibinfo{publisher}{USENIX Association},
  \bibinfo{address}{USA}, Article \bibinfo{articleno}{38},
  \bibinfo{numpages}{16}~pages.
\newblock
\showISBNx{978-1-939133-19-9}


\bibitem[Rychnovsk\'{y} and Holubov\'{a}.(2015)]%
        {Rychnovsky15}
\bibfield{author}{\bibinfo{person}{Du\v{s}an Rychnovsk\'{y}} {and}
  \bibinfo{person}{Holubov\'{a}.}} \bibinfo{year}{2015}\natexlab{}.
\newblock \showarticletitle{Generating XML Data for XPath Queries.}
\newblock \bibinfo{journal}{\emph{Association for Computing Machinery.}}
  (\bibinfo{year}{2015}).
\newblock
\urldef\tempurl%
\url{https://doi.org/10.1145/2695664.2695691}
\showDOI{\tempurl}


\bibitem[Slutz(1998)]%
        {slutz1998massive}
\bibfield{author}{\bibinfo{person}{Donald~R. Slutz}.}
  \bibinfo{year}{1998}\natexlab{}.
\newblock \showarticletitle{Massive Stochastic Testing of SQL}. In
  \bibinfo{booktitle}{\emph{Proceedings of the 24rd International Conference on
  Very Large Data Bases}} \emph{(\bibinfo{series}{VLDB '98})}.
  \bibinfo{publisher}{Morgan Kaufmann Publishers Inc.}, \bibinfo{address}{San
  Francisco, CA, USA}, \bibinfo{pages}{618–622}.
\newblock
\showISBNx{1558605665}


\bibitem[Sotiropoulos et~al\mbox{.}(2021)]%
        {Sotiropoulos21}
\bibfield{author}{\bibinfo{person}{Thodoris Sotiropoulos},
  \bibinfo{person}{Stefanos Chaliasos}, \bibinfo{person}{Vaggelis Atlidakis},
  \bibinfo{person}{Dimitris Mitropoulos}, {and} \bibinfo{person}{Diomidis
  Spinellis}.} \bibinfo{year}{2021}\natexlab{}.
\newblock \showarticletitle{Data-Oriented Differential Testing of
  Object-Relational Mapping Systems}. In \bibinfo{booktitle}{\emph{2021
  IEEE/ACM 43rd International Conference on Software Engineering (ICSE)}}.
  \bibinfo{pages}{1535--1547}.
\newblock
\urldef\tempurl%
\url{https://doi.org/10.1109/ICSE43902.2021.00137}
\showDOI{\tempurl}


\bibitem[Su et~al\mbox{.}(2021)]%
        {SuTing2021}
\bibfield{author}{\bibinfo{person}{Ting Su}, \bibinfo{person}{Yichen Yan},
  \bibinfo{person}{Jue Wang}, \bibinfo{person}{Jingling Sun},
  \bibinfo{person}{Yiheng Xiong}, \bibinfo{person}{Geguang Pu},
  \bibinfo{person}{Ke Wang}, {and} \bibinfo{person}{Zhendong Su}.}
  \bibinfo{year}{2021}\natexlab{}.
\newblock \showarticletitle{Fully Automated Functional Fuzzing of Android Apps
  for Detecting Non-Crashing Logic Bugs}.
\newblock \bibinfo{journal}{\emph{Proc. ACM Program. Lang.}}
  \bibinfo{volume}{5}, \bibinfo{number}{OOPSLA}, Article
  \bibinfo{articleno}{156} (\bibinfo{date}{oct} \bibinfo{year}{2021}),
  \bibinfo{numpages}{31}~pages.
\newblock
\urldef\tempurl%
\url{https://doi.org/10.1145/3485533}
\showDOI{\tempurl}


\bibitem[Todic and Uzelac(2012)]%
        {Todic12}
\bibfield{author}{\bibinfo{person}{Milos Todic} {and}
  \bibinfo{person}{Branislav Uzelac}.} \bibinfo{year}{2012}\natexlab{}.
\newblock \showarticletitle{Combined XML/XQuery generator}.
\newblock \bibinfo{journal}{\emph{Proceedings of the Fifth International
  Workshop on Testing Database Systems}} (\bibinfo{year}{2012}).
\newblock
\urldef\tempurl%
\url{https://doi.org/10.1145/2304510.2304519}
\showDOI{\tempurl}


\bibitem[Wu et~al\mbox{.}(2009)]%
        {Wu09}
\bibfield{author}{\bibinfo{person}{Yuqing Wu}, \bibinfo{person}{Namrata Lele},
  \bibinfo{person}{Rashmi Aroskar}, \bibinfo{person}{Sharanya Chinnusamy},
  {and} \bibinfo{person}{Sofia Brenes}.} \bibinfo{year}{2009}\natexlab{}.
\newblock \showarticletitle{XQGen: An Algebra-Based XPath Query Generator for
  Micro-Benchmarking}. In \bibinfo{booktitle}{\emph{Proceedings of the 18th ACM
  Conference on Information and Knowledge Management}} (Hong Kong, China)
  \emph{(\bibinfo{series}{CIKM '09})}. \bibinfo{publisher}{Association for
  Computing Machinery}, \bibinfo{address}{New York, NY, USA},
  \bibinfo{pages}{2109–2110}.
\newblock
\showISBNx{9781605585123}
\urldef\tempurl%
\url{https://doi.org/10.1145/1645953.1646328}
\showDOI{\tempurl}


\bibitem[Yang et~al\mbox{.}(2011)]%
        {yang2011finding}
\bibfield{author}{\bibinfo{person}{Xuejun Yang}, \bibinfo{person}{Yang Chen},
  \bibinfo{person}{Eric Eide}, {and} \bibinfo{person}{John Regehr}.}
  \bibinfo{year}{2011}\natexlab{}.
\newblock \showarticletitle{Finding and Understanding Bugs in C Compilers}.
  \bibinfo{publisher}{Association for Computing Machinery},
  \bibinfo{address}{New York, NY, USA}.
\newblock
\showISBNx{9781450306638}
\urldef\tempurl%
\url{https://doi.org/10.1145/1993498.1993532}
\showURL{%
\tempurl}


\bibitem[Z et~al\mbox{.}(2023)]%
        {Hua23}
\bibfield{author}{\bibinfo{person}{Hua Z}, \bibinfo{person}{Lin W},
  \bibinfo{person}{Ren L}, \bibinfo{person}{Li Z}, \bibinfo{person}{Zhang L},
  \bibinfo{person}{Jiao W}, {and} \bibinfo{person}{Xie T.}}
  \bibinfo{year}{2023}\natexlab{}.
\newblock \showarticletitle{GDsmith: Detecting bugs in Cypher graph database
  engines.}
\newblock \bibinfo{journal}{\emph{Proceedings of ACM SIGSOFT International
  Symposium on Software Testing and Analysis}} (\bibinfo{year}{2023}).
\newblock
\urldef\tempurl%
\url{https://doi.org/10.48550/arXiv.2206.08530}
\showDOI{\tempurl}


\bibitem[Zhang et~al\mbox{.}(2017)]%
        {zhang2017skeletal}
\bibfield{author}{\bibinfo{person}{Qirun Zhang}, \bibinfo{person}{Chengnian
  Sun}, {and} \bibinfo{person}{Zhendong Su}.} \bibinfo{year}{2017}\natexlab{}.
\newblock \showarticletitle{Skeletal Program Enumeration for Rigorous Compiler
  Testing}. In \bibinfo{booktitle}{\emph{Proceedings of the 38th ACM SIGPLAN
  Conference on Programming Language Design and Implementation}} (Barcelona,
  Spain) \emph{(\bibinfo{series}{PLDI 2017})}. \bibinfo{publisher}{Association
  for Computing Machinery}, \bibinfo{address}{New York, NY, USA},
  \bibinfo{pages}{347–361}.
\newblock
\showISBNx{9781450349888}
\urldef\tempurl%
\url{https://doi.org/10.1145/3062341.3062379}
\showDOI{\tempurl}


\bibitem[Zheng et~al\mbox{.}(2022)]%
        {Zheng22}
\bibfield{author}{\bibinfo{person}{Yingying Zheng}, \bibinfo{person}{Wensheng
  Dou}, \bibinfo{person}{Yicheng Wang}, \bibinfo{person}{Zheng Qin},
  \bibinfo{person}{Lei Tang}, \bibinfo{person}{Yu Gao}, \bibinfo{person}{Dong
  Wang}, \bibinfo{person}{Wei Wang}, {and} \bibinfo{person}{Jun Wei}.}
  \bibinfo{year}{2022}\natexlab{}.
\newblock \showarticletitle{Finding bugs in Gremlin-based graph database
  systems via randomized differential testing.}
\newblock \bibinfo{journal}{\emph{Proceedings of the 31st ACM SIGSOFT
  International Symposium on Software Testing and Analysis}}
  (\bibinfo{year}{2022}).
\newblock
\urldef\tempurl%
\url{https://doi.org/10.1145/3533767.3534409}
\showDOI{\tempurl}


\bibitem[Zhong et~al\mbox{.}(2020)]%
        {zhong2020squirrel}
\bibfield{author}{\bibinfo{person}{Rui Zhong}, \bibinfo{person}{Yongheng Chen},
  \bibinfo{person}{Hong Hu}, \bibinfo{person}{Hangfan Zhang},
  \bibinfo{person}{Wenke Lee}, {and} \bibinfo{person}{Dinghao Wu}.}
  \bibinfo{year}{2020}\natexlab{}.
\newblock \showarticletitle{Squirrel: Testing database management systems with
  language validity and coverage feedback}. In
  \bibinfo{booktitle}{\emph{Proceedings of the 2020 ACM SIGSAC Conference on
  Computer and Communications Security}}. \bibinfo{pages}{955--970}.
\newblock


\end{thebibliography}

\end{document}